\documentclass[aps,nofootinbib,twocolumn,floatfix,showpacs,preprintnumbers,prd]{revtex4}
\usepackage{graphicx, color}
\usepackage[bookmarks=false]{hyperref}

\newcommand{\be}[1]{\begin{eqnarray}}
\newcommand{\ee}{\end{eqnarray}}

\newcommand{\lsi}{\,\raisebox{-0.13cm}{$\stackrel{\textstyle<}{\textstyle\sim}$}\,}
\newcommand{\gsi}{\,\raisebox{-0.13cm}{$\stackrel{\textstyle> }{\textstyle\sim}$}\,}

\begin{document}

\title{Study of the Gamma-ray Spectrum from the Galactic Center in view of Multi-TeV Dark Matter Candidates}

\author{Alexander V. Belikov}
\affiliation{Institut d'Astrophysique de Paris, UMR 7095 CNRS, Universit\'e Pierre et Marie Curie, 98 bis Boulevard Arago, Paris 75014, France}

\author{Gabrijela Zaharijas}
\affiliation{Abdus Salam International Centre for Theoretical Physics, Strada Costiera 11, 34151, Trieste, Italy}
\affiliation{Institut d'Astrophysique de Paris, UMR 7095 CNRS, Universit\'e Pierre et Marie Curie, 98 bis Boulevard Arago, Paris 75014, France}

\author{Joseph Silk}
\affiliation{Institut d'Astrophysique de Paris, UMR 7095 CNRS, Universit\'e Pierre et Marie Curie, 98 bis Boulevard Arago, Paris 75014, France}
\affiliation{Beecroft Institute of Particle Astrophysics and Cosmology, 1 Keble Road, University of Oxford, Oxford OX1 3RH UK}
\affiliation{Department of Physics and Astronomy, 3701 San Martin Drive, The Johns Hopkins University, Baltimore MD 21218, USA}

\date{\today}

\begin{abstract}

Motivated by the complex gamma-ray spectrum of the Galactic Center source now  measured 
over five decades in energy, we revisit the issue of the role of dark matter annihilations in this interesting region. We reassess whether the emission measured by the HESS collaboration could be a signature of
dark matter annihilation, and we use the {\em Fermi} LAT spectrum to model the
emission from  SgrA*, using power-law spectral fits. We find that good fits are achieved by a power law with an index $\sim 2.5-2.6$, in combination with a spectrum similar to the one  observed from pulsar population and with a spectrum from a $\gsi10$ TeV DM annihilating to a mixture of $b{\bar b}$ and harder $\tau^+ \tau^-$ channels and with boost factors of the order of a hundred.  Alternatively, we also consider the combination of a log-parabola fit with the DM contribution.
Finally, as both the spectrum of gamma rays from the Galactic Center and the spectrum of cosmic ray electrons exhibit a cutoff at TeV energies, we study the dark matter fits to both data-sets. Constraining the spectral shape of the purported dark matter signal provides a robust way of comparing data.  We find a marginal overlap only between the 99.999\% C.L. regions in parameter space.
\end{abstract}


\maketitle

\section{Introduction}

In the last few years, a number of new generation cosmic-ray detectors have provided high quality measurements of gamma rays ({\it Fermi} LAT \cite{fermi}, VERITAS \cite{veritas}, HESS \cite{hess}, MAGIC \cite{magic}), charged cosmic rays (PAMELA \cite{pamela}, CREAM \cite{cream}, {\it Fermi} LAT, HESS) and neutrinos (ICE CUBE \cite{icecube}, Antares \cite{antares}) at GeV and/or TeV energies. The newly acquired data are currently revolutionizing our understanding of the origins and propagation of Galactic cosmic rays and they are equally important for indirect searches for dark matter (DM). The theory of cold dark matter is supported by numerous pieces of evidence (e.g. an impressive agreement between predictions of N-body simulations and the observed large-scale structure of the Universe, Big-Bang Nucleosynthesis, gravitational lensing). Weakly Interacting Massive Particles (WIMPs) are dark matter candidates with the required cosmological properties and are naturally present in many beyond-the-Standard-Model theories, as a new particle of mass close to the Electro Weak (EW) breaking scale. An additional appeal of this paradigm is the so-called {\it WIMP miracle}: a TeV particle weakly coupled to the Standard Model (SM) particles that decouples thermally in the Early Universe, naturally has an abundance comparable to the DM abundance measured today. If it were to decay or annihilate to Standard Model particles, such as photons and electrons/positrons, the bulk of the emitted energy is expected to fall just below the DM mass, and to show features in this energy range.

The trajectory of gamma rays, unlike that of charged cosmic rays, is unaffected by the Galaxy, allowing the analysis of individual sources. In order to maximize the luminosity of gamma rays produced in DM annihilations,  a standard search strategy is to look at nearby sources with high DM density. An example of such 'good target' is the center of our Galaxy. 
It is one of the closest targets and it is expected to contain a high dark matter density, due to a possible cusp in the density  profile predicted by N-body simulations \cite{Navarro:1995iw,Graham:2005xx,Navarro:2008kc}. However, the dynamical center of our Galaxy hosts a supermassive black hole (compact radio source SgrA*) \cite{Melia:2001dy}, whose spectrum is measured over a range of wavelengths, spanning  radio and sub-mm \cite{Narayan:1997ku}, near-infrared \cite{Genzel:2003as,Clenet:2003tv}, X rays \cite{Baganoff:2001ju,Xu:2005et}, and $\gamma$ rays \cite{MayerHasselwander:1998hg,Kosack:2004ri,Tsuchiya:2004wv,Aharonian:2004wa,Aharonian:2009zk,Albert:2005kh,Beilicke:2011rn}, and which presents a complex background for DM searches \cite{Zaharijas:2006qb}.

The Very High Energy (VHE) gamma-ray source at the Galactic Center (GC), HESS J1745-290, was detected by the HESS collaboration in 2003 and observed since \cite{Aharonian:2004wa,Aharonian:2009zk}, in the range 200 GeV through 55 TeV (an early detection in VHE gamma rays was done by the WHIPPLE \cite{Kosack:2004ri} and CANGAROO-II \cite{Tsuchiya:2004wv} telescopes).  The GC is also observed by MAGIC \cite{Albert:2005kh} and VERITAS \cite{Beilicke:2011rn} with spectra compatible to the one measured by HESS. The large zenith angles for these experiments however, result in significantly smaller exposures. The gamma ray emission from the GC was subsequently interpreted both in terms of standard astrophysical emission originating in the vicinity of the  supermassive black hole \cite{Aharonian:2004jr,Aharonian:2005ti,Liu:2006bf,Wang:2009bv,Atoyan:2004ix} as in terms of a signal from self-annihilating dark matter \cite{Hooper:2004vp,Profumo:2005xd,Aharonian:2006wh,Cembranos:2012nj}. 

The {\it Fermi} satellite, launched in 2008, measured the emission from the GC in the lower energy range 100 MeV-100 GeV. In the first year {\it Fermi} LAT catalog 1FGL \cite{Abdo:2010ru}, the source closest to the GC, was classified as 1FGL J1745.6-2900c. While the most recent data from this complicated region have not yet been analyzed by the collaboration (however see {\it Fermi} LAT proceedings and talks \cite{Vitale:2009hr,JCT,CM}), a dedicated analysis has been done by several authors \cite{Chernyakova:2010ey,Boyarsky:2010dr,Hooper:2010mq,Hooper:2011ti}.  In these works it was found that the {\it Fermi} and HESS sources are spatially coincident (in agreement with the preliminary findings of the {\it Fermi} LAT team), though the issue of the extension of the GeV source is still unsettled.

At energies above 10 GeV, the {\it Fermi} LAT has a good angular resolution comparable to that of  HESS at TeV energies. This motivates the joint analysis of the two data-sets which, taken together, show the possible presence of one or more spectral breaks. In \cite{Chernyakova:2011zz}, it was shown that the spectra is well modeled by two  plateaux with a steeply falling region in between, and astrophysical mechanisms which rely on modeling collisions of energetic protons, accelerated by the supermassive black hole, with the surrounding gas have been proposed in order to match this spectral feature (for an earlier model which accommodates a break between the GeV and TeV photons originating in the vicinity of the SgrA* see \cite{Aharonian:2004jr}).  

Motivated by the complex gamma ray spectrum of the Galactic Centre source we revisit the issue of a multi-TeV dark matter fit to this region. We show that a single power law over the GeV and TeV range, with two bump-like features, provides a comparably good fit. We assume the feature at high-energy as a signature of DM annihilation. We then notice that the lower energy bump is curiously located around $\sim 3$ GeV, an average energy at which a harder emission from the pulsar population has a high energy cut-off \cite{Abdo:2009ax,Abazajian:2010zy}. In addition, we also show that a log-parabola in the low energy range (typical for spectra of some AGNs \cite{Collaboration:2011bm}) provides the same conclusions.

In other words,  we go back to the question of whether the emission measured by the HESS collaboration could explained by DM annihilation, but we use the {\it Fermi} LAT spectrum to model the standard astrophysical emission from the SgrA*\footnote{For work along similar lines, which instead uses the HESS data to model the standard astrophysical signal in the GeV range see \cite{Zaharijas:2006qb,Jeltema:2008hf}.}. 

It has been known from early works \cite{Hooper:2004vp,Profumo:2005xd,Aharonian:2006wh} that high energy HESS data, if interpreted as DM signatures, could be due to $\gsi$10 TeV DM, which is somewhat heavier than the standard expectation for a WIMP, and that the flatness of the HESS spectrum over two decades in energy is hard to fit with the usual WIMP spectrum. The second issue is somewhat alleviated in this work by the addition of a power law constrained by the {\it Fermi} LAT data. 

In this work, we focus on spectral features of the central source and leave morphological studies aside (for recent work addressing this issue see \cite{Linden:2012iv,Linden:2012bp}), which is expected to be a good approximation for strongly cusped DM profiles \cite{Serpico:2008ga,Regis:2008ij}. While there are indications that the GC source at GeV energies can be spatially extended \cite{Hooper:2010mq,Boyarsky:2010dr,Hooper:2011ti}, this issue is still unsettled due to the complexity of this active region and we do not address it here\footnote{Note that at TeV energies a more extended region, $\sim 2$ deg (200 pc) around the GC, known as the Galactic Centre Ridge \cite{Aharonian:2006auRIDGE} also has a puzzling spectrum, harder than the surrounding medium. This emission correlates spatially with a complex of giant molecular clouds in the central 200 pc of the Milky Way and one possible explanation, based on energetics of the emission, is that it could have come from a single supernova explosion injecting hard cosmic rays, around $10^4$ years ago.}. 

On the side of spectral signatures of heavy ($\gsi 100$ GeV) dark matter candidates, it has been recently noted that addition of the emission of W or Z bosons from final and internal states, neglected in standard computations, could substantially alter the spectrum \cite{Kachelriess:2009zy,Ciafaloni:2010ti,Bell:2011eu,Ciafaloni:2011sa,Bell:2011if,Ciafaloni:2011gv}, both at low energies and in the energy range close to the DM mass, which is critical for the spectral fits in the GC region. As the magnitude and shape of the spectral corrections in the hard energy region are model dependent we do not consider them here. We however comment on the implications of such corrections on our fits in section \ref{sec:spectrum}.

Recently there have been strong indications of a line signature in the {\it Fermi} LAT data at around 130 GeV, with a likely origin within $\lsi 4^\circ$ within the GC \cite{Weniger:2012tx,Boyarsky:2012ca,Tempel:2012ey}. The morphology of this emission was studied in \cite{Su:2012ft} finding a best fit position $\sim 1.5^\circ$ off the Galactic Center, with a $\sim 5 \sigma$ significance of the signal compared to the null-hypothesis. As the nature of this signature is still not clear at the moment, we do not address it here.  We simply notice that if the signal proves to be of a non-instrumental origin a DM interpretation would be well motivated (but see \cite{Buchmuller:2012rc,tracyline} for a difficulty to reach needed branching ratios to line channels in the usual WIMP models and  \cite{Felixline} for alternative astrophysical explanation of such signal). 

In addition to the gamma-ray data analysis, we make a connection with another astrophysical measurement which recently has shown a surprising feature: the local population of electrons/positrons as measured by PAMELA, {\it Fermi} LAT and HESS experiments (see the recent review in \cite{Cirelli:2012tf} for both dark matter annihilation and astrophysical interpretations). While it has been shown already that dark matter is viable but not necessary to explain these data sets, we return here  to the DM interpretation in an attempt to do a more careful calculation for heavy candidates. We consider the full relativistic calculation of energy losses as implemented in the GALPROP code \cite{Porter:2008ve,Vladimirov:2010aq}, critical for such high energies, and, as opposed to the gamma-ray case, we account for electroweak bremsstrahlung contributions to the electron and positron spectra, which can affect positron fits performed at low energies. Na\" ively, as the high energy cut-off in the HESS gamma-ray data is higher than that measured by the HESS electron data, these data imply a different mass of the potential DM candidate. We nevertheless explore this possibility by deriving the best-fit DM regions, marginalizing over DM and astrophysical modeling. 
 
The article is organized as follows: in the section \ref{sec:astroGC} we briefly summarize observations on the GC source focusing on gamma-ray spectra; in section  \ref{sec:dist} we review DM density profiles in the central parts of our halo and annihilation spectra of heavy DM candidates in \ref{sec:spectrum}. In section \ref{sec:fitGC} we describe the fits we perform to the Galactic center gamma-ray data, while in section  \ref{sec:elpos} we focus on the electron and positron measurement. We compare our findings with the current constraints in the literature on multi-TeV DM models in \ref{sec:const} and summarize in section \ref{sec:summary}.

\section{Astrophysics at the Galactic Center} 
\label{sec:astroGC}

Below we briefly summarize current observations on the GC, and some of the proposed emission models in the gamma-ray range.

Emission from Sgr A* in the radio to sub-mm presents a very hard luminosity spectrum $L_\nu \sim \nu^\alpha$ with an index  $\alpha =0.8$ with a cut-off at about $\nu \sim$ 103 GHz \cite{Narayan:1997ku}. The angular size of the source depends on the frequency of observation: at 1 GHz, it is of the order of 1.5 arcsec while it reduces to 0.2 milliarcsec at 86 GHz. Variability (though without a clear pattern) is observed at these frequencies. 

The quiescent flux in the near-infrared has been detected with the VLT \cite{vlt}, in the energy range 0.3--1 eV, as a point source with a position coincident with the supermassive black hole within an accuracy of 10-20 marcses. Chandra X-ray observatory \cite{chandra}, covering the energy range 0.1--10 keV with an angular resolution of 0.5 arcsec, observed the GC as an extended source with a size of  1.5 arcsec. The near-infrared and X-ray emissions from Sgr A* are characterized by a large variability (on different timescales in the two cases). In addition to the central source, continuum X-ray and radio observations indicates the central 10 pc region of the Galaxy (comparable to the resolution of gamma-ray detectors) as a complex environment characterized by several distinct structures \cite{Melia:2001dy,Crocker:2010xc}.

In VHE gamma rays, the latest analysis of the 93 hours of live time observation of the central source by the HESS collaboration \cite{Aharonian:2009zk} revealed that the spectra of HESS J1745-290 cannot be fitted with a single power law, but that instead a power law spectrum with an exponential cut-off characterized by a photon index of $2.10~\pm~0.04~({\rm stat})~\pm~0.10~({\rm syst})$ and a cut-off energy at $15.7~\pm~3.4~({\rm stat})~\pm~2.5~({\rm syst})$ TeV, provides a good fit. A dedicated analysis in \cite{2010MNRAS.402.1877A}  was able to localize the position of the TeV point source within $7.3$ arcsec $\pm ~8.7$ arcsec (stat) $\pm ~8.5$ arcsec (syst) from Sgr A*, consistent as well with a pulsar wind nebula 8.7 arcsec from the GC \cite{Wang:2005ya}\footnote{An angular resolution of approximately 0.1 deg corresponds the distance of {15} pc, when Solar position at 8.5 kpc is assumed.}. Unlike in the radio and X-Ray observations of the GC, no variability has been observed in the high energy regime  \cite{Aharonian:2009zk}, potentially indicating that the $\gamma$-ray emission mechanism differs from the low energy regime.

The authors of \cite{Chernyakova:2011zz} analyzed 25 months of the {\it Fermi} LAT data using the P6V3 instrument response function, modeling the diffuse emission and point sources in the $10^{\circ}\times10^{\circ}$ region and in the 100 MeV--300 GeV energy range. The spectral curve for the central source 1FGL J1745.6-2900 was derived to be a power law with a slope $\Gamma$ = 2.212 $\pm$ 0.005 and a flux normalization F = (1.39 $\pm$ 0.02) $\times~10^{-8}$ cm$^{-2}$ s$^{-1}$ MeV$^{-1}$ at 100 MeV. The authors noted that a better fit is obtained when the spectrum is described by two distinct power laws. In this case, the fitted slope is equal to $\Gamma$ = 2.196$\pm$0.001 in 300 MeV\--5 GeV energy range, and $\Gamma$= 2.681$\pm$0.003 in 5\--100 GeV energy range\footnote{It has also been noted in \cite{Chernyakova:2011zz} that the error of the lowest data point might be underestimated. {\it Fermi} LAT has a very broad PSF at lower energies, starting from 4$^\circ$ at 100 MeV and rapidly improving to $0.1^\circ$ at $\gsi$ 10 GeV. Thus, taking into account the possible source confusion in the region, one should treat the first point in the spectrum (100 \-- 300 MeV) with caution.}.  It was also shown in this work that {\it Fermi} LAT source 1FGL J1745.6-2900 lies within the error box of HESS source J1745-290. 

Various models predict VHE emission from the central source, produced either close to the black hole itself \cite{Aharonian:2004jr, Atoyan:2004ix} or within an $\sim10$ pc zone around Sgr A* due to the interaction of run-away protons with the ambient medium \cite{Aharonian:2005ti,Liu:2006bf,Wang:2009bv}.

A theoretical model proposed in \cite{Chernyakova:2011zz} which can explain the shape of the combined {\it Fermi} LAT and the HESS spectrum considers the hadronic interactions of relativistic protons which, having diffused away from a central source, presumably the central black hole, fill the inner few parsecs of our Galaxy. The size of the region combined with the energy dependence of the diffusion then result in a specific spectral shape. Authors show that the spectra can be well fit  after  several fitting parameters are adjusted (free parameters in this model characterize geometry, diffusion and injection rate history). This model has the additional appeal of reconciling variability of  SgrA* at lower energies \cite{Terrier:2010bn} with no variability in gamma rays (as the gamma ray emission originates from a larger region surrounding the black hole and is emitted during the diffusion of the relativistic protons through the interstellar medium). However, due to our poor understanding of the spectrum  of protons accelerated by the central black hole, or of past activity of SgrA*, it is not simple to independently test the best-fit parameters in this model. 

In this work we remain agnostic about the astrophysical mechanism responsible for the emission of SgrA* and we sample different fitting functions to the {\it Fermi} LAT data in order to explore the uncertainty it brings into the DM interpretation of the HESS measurement.

\section{DM signal: distribution and spectra}

\subsection{Signal} \label{sec:dist}

The flux of gamma rays from annihilating dark matter in the direction $\psi$ can be written in the following way:
\be{l}
\Phi_\chi(E, \psi) = \frac{dN}{dE_\gamma} (E) \frac{\langle \sigma v \rangle}{8\pi M^2_\chi} \int_{los} \rho^2(r) dl\,,
\ee
where $M_\chi$ is the mass of the WIMP, $\langle \sigma v \rangle$ is the standard thermal cross-section (assumed to $3\times10^{-26}$ unless otherwise stated), $\rho(r)$ is the dark matter density as a function of distance $r = \sqrt{l^2 + D^2 - 2lD\cos{\psi}}$ away from the center. 
The generalized NFW density profile is given by $\rho(r) = \frac{\rho_0}{(r/a)^\gamma(1 + r/a)^{3-\gamma}}$, where $a$ is a length scale, $\rho_0$ is the density scale and $\gamma$ sets the inner slope.
The total flux for a source observed in a solid angle $\Delta\Omega$ is 
\be{lll}
\phi_\chi(E) &=& \int_{\Delta\Omega} d\Omega \:\Phi_\chi(E, \psi) \nonumber\\
& = &  \frac{dN}{dE_\gamma} (E) \frac{\langle \sigma v \rangle}{8\pi M^2_\chi}   \int_\Delta\Omega\int_{los} \rho^2(r) dl\,d\Omega \nonumber\\
& = & \Phi_\chi(E)\, J (\Delta\Omega)\,.
\ee
The particle physics and energy dependence is contained in the function $\Phi_\chi(E)$, while the astrophysics is contained in the factor $J$. We assume the spectrum $dN/dE_\gamma$ to be given by a sum over annihilation channels 
\be{}
\frac{dN}{dE_\gamma} = \sum_i \frac{dN}{dE_i}\,.
\ee
We assume that $\Delta\Omega = 10^{-5}$ sr (corresponding to the definition of a point source given the $0.1^\circ$ resolution of HESS). As we perform only spectral fits, the details of the profile shape are not critical. In what follows we assume the standard NFW dark matter density profile \cite{Navarro:1995iw} with the following normalization parameters: the Milky Way mass $M_{\rm\small MW} = 1.6\times 10^{12} M_{\odot}$ \cite{Gnedin:2010fv}, the dark matter density $\rho_{\odot} (R_\odot = 8.5\, \mbox{kpc}) = 0.3\,{\rm GeV}\,{\rm cm}^{-3}$ and inner slope  $\gamma = 1.0$. 

In the following we will express part of our results in terms of a 'boost factor' $B_F$. We define it as an enhancement of the DM induced gamma-ray signal over the one predicted based on the $J$ factors calculated for the NFW profile and assuming the value of the velocity averaged cross section of $\langle \sigma v \rangle_{\rm th} = 3\times 10^{-26}$ cm$^{3}$ s$^{-1}$. This value of the cross section leads to the correct relic abundance of DM in the simplest thermal decoupling scenarios and depends only logarithmically on the DM mass.  We note here that large boost factors could originate both from an enhancement in the cross section or from an enhanced DM density. Non-perturbative electro-weak resonant effects could naturally become relevant for heavy DM candidates and can enhance annihilation cross sections, however only for model-dependent narrow ranges of the DM mass, \cite{Hisano:2004ds}. We will mention some other possibilities in the following section.

The contraction of the dark matter density profile around the black hole residing in the center of the Galaxy could result in a so-called dark matter `spike', a region of enhanced density. It could form due to the adiabatic growth of the black hole, due to scattering of dark matter with the dense stellar environment around the black hole or due to baryonic infall \cite{Gondolo:1999ef, Gnedin:2003rj, Prada:2004pi} (however events such as binary black hole mergers or off-center supermassive black hole formation could results in the opposite effect.) In the simple adiabatic contraction model the slope of the spike is derived in terms of the inner slope $\gamma$ as $\gamma_{sp} = 2 + 1/({4-\gamma})$ and has an effect at distances below few parsecs from the black hole. Any of the above-mentioned mechanisms could provide an enhancement of the dark matter annihilation signal by a factor of 100 to 1,000 in the innermost region of the Galaxy (for comparison of $J$-factors in such scenarios we refer the reader to \cite {Bertone:2005hw}.)

\subsection{Spectrum} \label{sec:spectrum}
 
The spectrum of  multi-TeV DM particles has to be treated with care as in collisions with center-of mass energies significantly higher than the weak scale the radiation of electroweak gauge bosons from the products of the DM annihilation/decay becomes possible. The so-called {\it electroweak radiative corrections}, neglected until recently  \cite{Kachelriess:2009zy,Ciafaloni:2010ti,Bell:2011eu,Ciafaloni:2011sa,Bell:2011if,Ciafaloni:2011gv}, are expected to have a sizable impact on the energy spectra and branching ratios of SM particles originated from the annihilation/decay of DM particles with mass larger than the electroweak scale. 

Under the assumption that tree-level cross sections for processes leading to two final states are dominant over those corresponding to processes with three final states including a gauge boson it is then generically possible to compute such corrections in a model independent way \cite{Ciafaloni:2010ti}. These calculations have been implemented in the \texttt{PPPC4DMID} code \cite{Cirelli:2010xx}\footnote{This calculation assumes that final states are massless, which is in general a good approximation given the high center-of-mass energies that we deal with here.} which we use to to derive the spectra of gammas and electrons from DM annihilations. 

A specific consequence of radiative corrections, relevant to our case, is that the total energy is distributed over a large number and variety of low energy particles amplifying the spectrum at lower energies ($E\ll m_{\chi}$).  We therefore do not expect them to impact our fits of gamma rays and electrons which are measured at high energies, close to the best-fit particle mass. However, positrons are measured at lower energies, where EW corrections affect the spectrum by up to an order of magnitude \cite{Kachelriess:2009zy,Ciafaloni:2010ti,Bell:2011eu,Ciafaloni:2011sa,Bell:2011if,Ciafaloni:2011gv}. 

Nevertheless, it should be mentioned that in some specific cases radiative corrections can lead to large, model-dependent effects on the spectrum at high energies, with hard contributions near the DM mass. This can happen for the emission of photons radiated from virtual charged particles and are particular important for Majorana DM or `wino-like' scenarios \cite{Bergstrom:1989jr,Flores:1989ru,Bergstrom:2008gr}. These effects lead to the appearance of a sharp line-like signature, which could significantly increase the sensitivity of the DM searches \cite{Bringmann:2011ye,Bringmann:2012vr}. However, the inclusion of these channels does not significantly improve the quality of our gamma-ray fits (as also found in \cite{Profumo:2005xd}). Recent works \cite{Ciafaloni:2011sa,Ciafaloni:2011gv,Ciafaloni:2012gs} generalized the three-body final states calculations to also include the emission of EW bosons. These however affect the gamma-ray spectrum in the lower $x=E_{\gamma}/m_{DM}\lsi ~0.3$ region and are typically rather flat in energy \cite{Ciafaloni:2011sa} in this range. As the exact spectra and amplitude of these effects are model-dependent we do not consider them here.

In \cite{Ciafaloni:2012gs} it was shown that for the `wino-like' DM candidates which are expected to mostly produce $WW$ states via inclusion of photon emissions from the initial states actually can make comparable annihilations  to three-body final states including fermions $ffV$ (depending on the choice of couplings and exchange particle mass). This is interesting in the view of our analysis here as, as we will see, the gamma ray spectrum is best fit with a combination of $WW$ and harder $\tau \tau$ channels. This issue merits a dedicated investigation which is beyond the scope of this work.

In our analysis, the sum runs over $\mu^+\mu^-$, $\tau^+\tau^-$, $b\bar b$ and $t\bar t$ annihilation channels. As the DM annihilations to the $b\bar b$ channel result in a gamma-ray spectrum which is very similar to the spectrum of $t\bar t$ and $W^+ W^-$ we will use it to mimic all non-leptonic annihilation channels. 

\begin{center}
{\begin{table*}[ht]
\hfill{}
\begin{tabular}{c|c|c|c|c|c|c|c|c}
\hline
case & datasets & $\Gamma$ & A  $\times 10^{-8}$& GeV range fit & DM mass [TeV] & ($Br_{\tau^+\tau^-}$; $Br_{b\bar b}$) & $ B^{(2)}_{NFW}$ & $\chi^2$ /d.o.f.  \\ 
\hline
\hline
G.1 & {\it Fermi}+HESS				& 2.48 & $6.45 $ &	-		&	-			& -			&	-				&	5.48	\\
G.2 & {\it Fermi}+HESS				& 2.51 & $6.63$ &	-		&	12.7	& 	(1.0;0.0)		&		390		& 3.34 \\
\hline
G.3 & {\it Fermi}+HESS				& - & $0.25$ 	&	$(E/4.2)^{-2.54-0.07 {\rm log}(E/4.2)}$	&	15.5	& (0.45;0.55)		&		1645	& 1.4 \\
G.4 & {\it Fermi}+HESS				& 2.48 & $4.8$ 	&	$E^{-1.5}{\rm e}^{-E/3.4}$		&	12.2	& (0.8;0.2)		&		507 	& 1.0\\
G.5 & {\it Fermi}+HESS				& 2.55 & $4.4$ 	&	$m_{\chi}=28,~B_F=62$	&	18.2	& (0.48;0.52)		&		1400 	& 0.9\\
\hline
G.6 & HESS							& 2.51 & $3.21$ 	&	-		&	18.2	& (0.53;0.47)	&		1075 	& 0.81\\
G.7 & {\it Fermi}+HESS(full)	& 2.59 & $3.54$ 	&	$E^{-1.44}{\rm e}^{-E/3.8}$ &	14.9	& (0.47;0.53)	&		1322 	& 1.42\\
G.8 & {\it Fermi}+HESS(full)	& 2.51 & $3.98$ 	&	28	&	18.2	& (0.42;0.58)		&		1685 	& 1.09\\
\hline
\hline
\end{tabular}
\hfill{}
\caption{Best-fit parameters and $\chi^2$ for different models of the galactic center gamma-ray spectra.\label{tabGC}}
\end{table*}}
\end{center}

\section{Fitting gamma rays from the galactic center}
\label{sec:fitGC}

In this work we use the spectral analysis of {\it Fermi} LAT data from \cite{Chernyakova:2010ey} and HESS data from \cite{Aharonian:2009zk}. We study them in terms of an astrophysical component and a dark matter particle annihilation spectrum, to a combination of representative channels ($\mu^+\mu^-$, $\tau^+\tau^-$, $b\bar b$). Specifically, under the assumption that a power-law spectrum provides a reasonable fit to point sources (see, e.g., the analysis of the 1FGL catalog, \cite{Collaboration:2011bm}), our model for the joint HE and VHE spectra of the central source (1FGL J1745.6-2900) consists of the combination of a power-law and two additional contributions. 

As a possible alternative, we consider as well a  `log-parabola' spectrum (shown to better reproduce the spectrum of AGNs in the 2FGL) which adds two parameters while decreasing smoothly at high energy. This will be given by
\be
dN/dE=K(E/E_0)^{-\alpha-\beta log(E/E_0)}\,,
\ee
with parameters $K$, $\alpha$ (spectral slope at $E_0$) and the curvature $\beta$, while $E_0$ is an arbitrary reference energy.

In analyzing the data,  we consider three different sets of data. The first (A) consists of the {\it Fermi} LAT data excluding the last data point, assumed to be an outlier, and the HESS data, excluding the last four data points. These omissions are justified by their large error-bars and the potential difficulty in distinguishing the signal from the diffuse emission from the larger ROI in the more uncertain high energy region. In the second data set (B), we consider just the HESS data, still excluding the last four data points. The third case (C) is given instead by the full HESS and {\it Fermi} LAT data sets. 

The choice of these different combinations is motivated by our specific interest in multi-TeV cutoff apparent in the HESS data. Since the best-fit model would, to a large extent be determined by the {\it Fermi} LAT data which present much smaller errors, a comparison of the results from sets A and B is particularly relevant. In what follows we will be mostly considering set A, unless otherwise specified.

As mentioned above, we fit different functional forms to mimic possible astrophysical contributions in the GeV energy range and we add a DM spectrum to model the VHE range. In each case, we minimize the $\chi^2$ over the mass of  the dark matter candidate as well as the normalization of the DM flux, and the parameters of the additional function(s). We also marginalize over the branching ratios to $\mu^+\mu^-$, $\tau^+\tau^-$ and $b\,\bar b$ (with this last spectrum mimicking all non-leptonic annihilation channels).

\begin{figure*}[h]
\begin{tabular}{cc}
Cases G.1 and G.2 (data set A) & Case G.3 (data set A) \\
& \\
\includegraphics[width=0.49\textwidth]{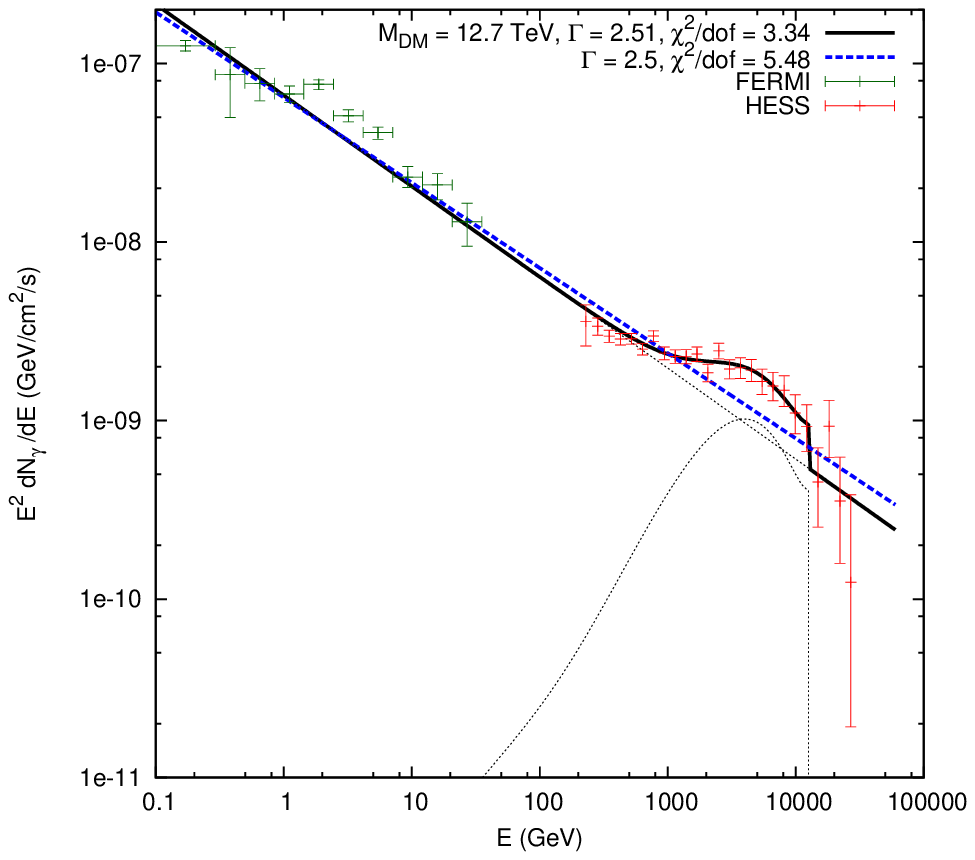} &
\includegraphics[width=0.49\textwidth]{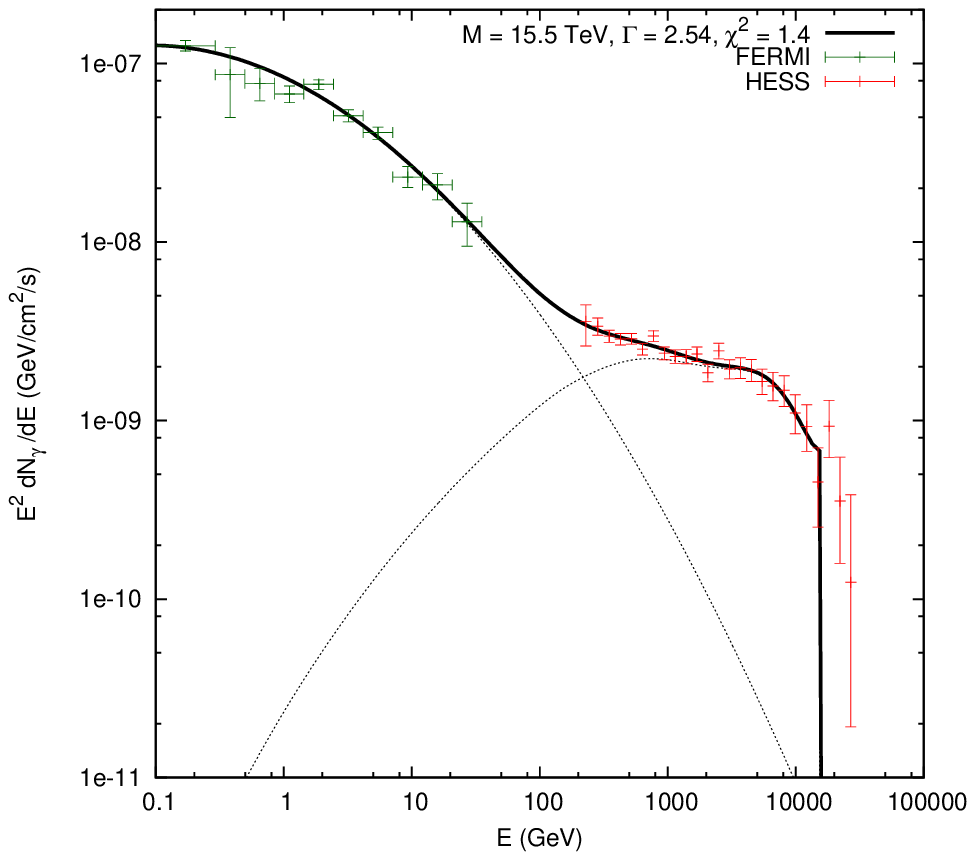} \\
& \\
Cases G.4 and G.5 (data set A) & Cases G.7 and G.8 (data set C) \\
& \\
\includegraphics[width=0.49\textwidth]{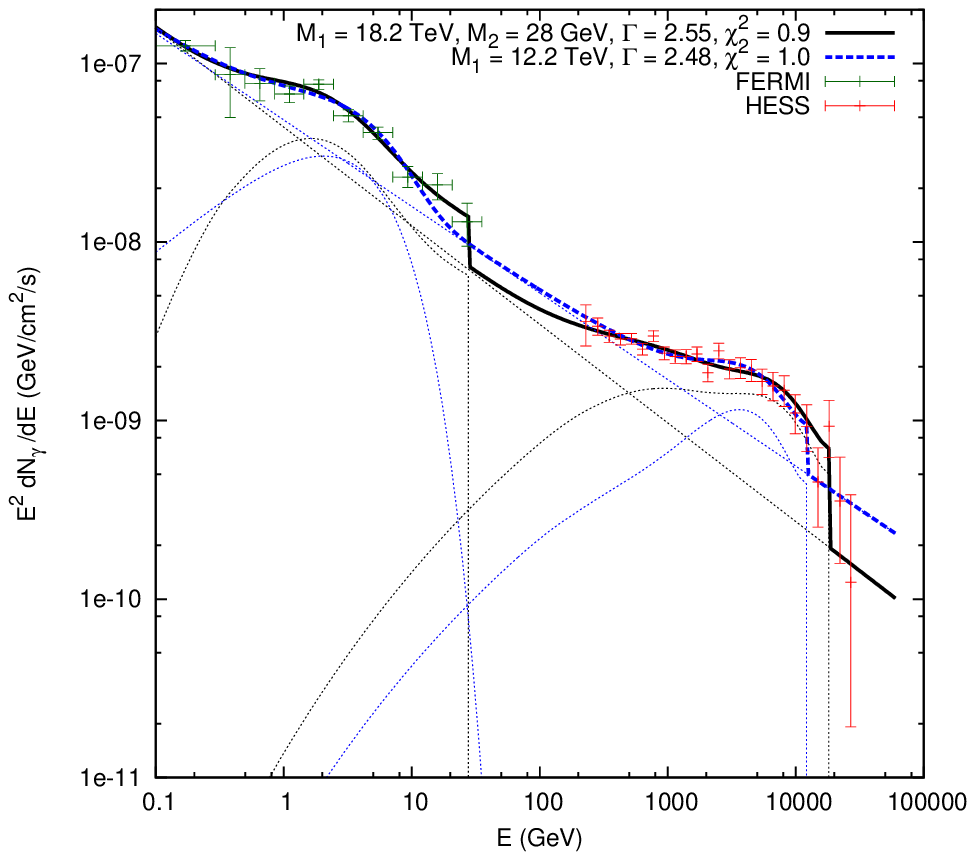} &
\includegraphics[width=0.49\textwidth]{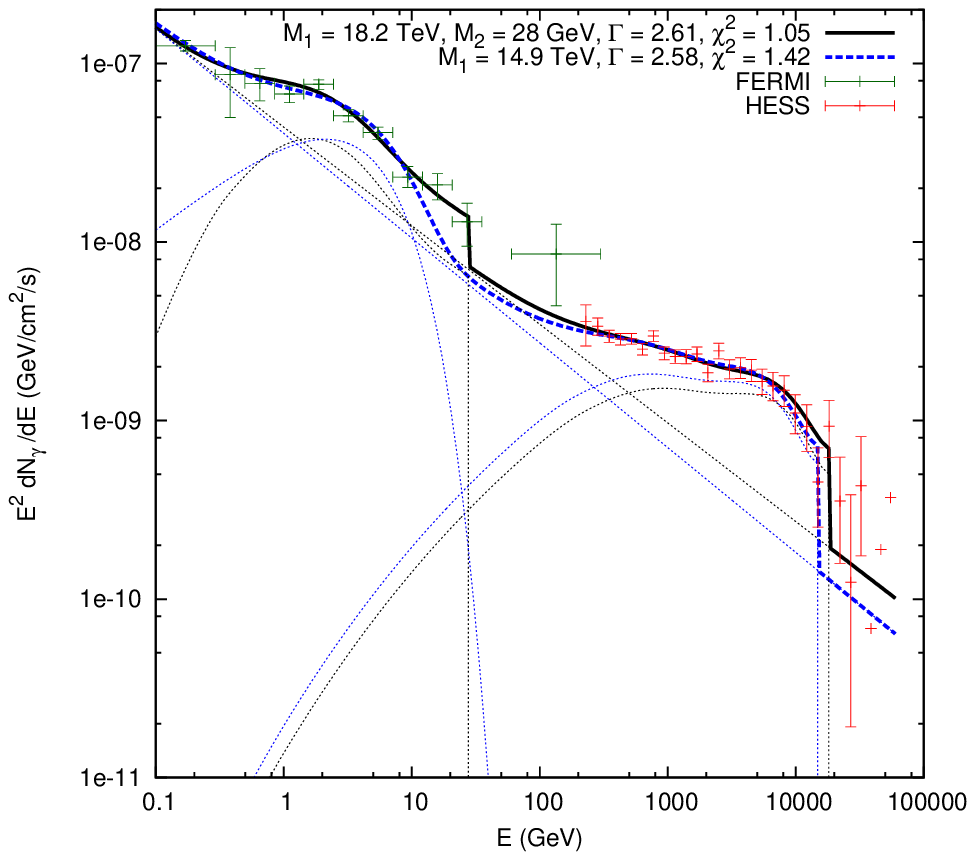} 
\end{tabular}
\caption{Fits to the gamma-ray spectrum of the source 1FGL J1745.6-2900 using {\it Fermi} LAT and HESS data. {\it Top, left panel}: single power law, case G.1 ({\it dashed, blue curve}) and a power law in combination with a heavy dark matter annihilation spectrum, case G.2 ({\it solid, black curve}). {\it Top, right  panel}: log-parabola spectrum, modeling an average pulsar, and a heavy dark matter annihilation spectrum, case G.3 ({\it solid curve}). {\it Bottom, left panel}: a power-law and two dark annihilation spectra, case G.5 ({\it solid, black curve}) and a power law with an exponential cutoff, modeling the pulsar contribution, case G.4 ({\it dashed, blue curve}). {\it Bottom, right panel}: same the bottom left panel but for data set C, cases G.7 ({\it dashed, blue curve}) and G.8 ({\it solid, black curve}). In all panels , the dark matter annihilation spectra, the power laws and the exponential cutoff function are plotted separately as dotted lines. See Table~\ref{tabGC} for details.}
\label{figSpecGC}
\end{figure*}

We start by fitting a power-law of the form $Ae^{-\Gamma}$ alone (case G.1) and a combination of a power law with the spectrum produced by annihilating dark matter $Ae^{-\Gamma} + \phi_{\chi}(E)$ (case G.2) to the data set A, see Fig. \ref{figSpecGC}, top left panel. These two cases clearly cannot accurately reproduce the data sets, and we quote values of $\chi^2$ in Table \ref{tabGC} just for reference. In the following we try to improve the low energy fit in order to gauge the impact on the DM fits in the high energy range.

In the first place, we model the emission in the GeV region with an addition of a power law with an exponential cut-off to mimic the  potential contribution of an unresolved pulsar population in this region. This will be given by 
\be{}
E_{\rm puls}=E^{-\gamma}{\rm e}^{-E/E_{cut}}\,.
\ee
In this case we find that the best-fit parameters are $E_{cut}=3.8$ GeV, $\gamma=1.6$, intriguingly similar to the mean values found in the {\it Fermi} LAT pulsar catalog \cite{Abdo:2009ax}. We do not attempt to make strong statements about the origin of this feature, as it might be challenging  to localize a significant pulsar population in a tiny region of about $30$ pc\footnote{corresponding to the PSF of {\it Fermi} LAT at 3 GeV which is $\gsi 0.2$ deg} (pulsar kicks of 100 km/s allow them to move 10 pc over 0.1 Myr time, assuming they are not gravitationally bound), but we note that that such a spectrum provides a good fit in this case. An HST-based measurement performed in  \cite{Figer:2003tu} suggests a high star formation rate in this region rich with young (few Myrs) stellar clusters which would naturally produce a numerous pulsar population. In addition older clusters could spiral into the central parsec region drawn by dynamical friction \cite{2001ApJ...546L..39G} (see also \cite{Abazajian:2010zy} for an interpretation of the extended source SgrA* by millisecond pulsars). We denote the model that includes this exponential cutoff to the power-law and the annihilation spectrum of heavy dark matter fit to data set A as case G.4 in Table \ref{tabGC} (see Fig. \ref{figSpecGC}, bottom left panel). In addition, and following the {\it Fermi} LAT catalog 2FGL, we model the GeV emission by the log-parabola fit, case G.3. We find the best-fit values to be $\alpha = 2.5$, $\beta = 0.074$ and $E_0 = 4.2$ GeV. This case is shown in the top right panel of Fig.~\ref{figSpecGC} .

Lastly, we model the low energy spectrum with a power law and an additional DM spectrum, which we again marginalize over all the DM masses and channels, case G.5 (Fig. \ref{figSpecGC}, bottom left panel). In this way, we allow the most freedom to the fit, and this check is meant to test whether there is a significant impact on the heavy DM derived regions, by allowing for a large freedom in the fits to the low energy data. The $\chi^2$ values for these choices are comparable as well as are the  inferred properties of the heavy DM.

In case G.8 we consider the full {\it Fermi} and HESS datasets C (Fig. \ref{figSpecGC}, bottom right panel) and check that the fit parameters are not too different from the ones derived in the analysis of dataset A. 

As an additional check we consider as well an analysis limited to the HESS data (data set B), defined by a power law and a heavy dark matter annihilation spectrum (case G.6 in Table~\ref{tabGC}). As mentioned before, a power law with an exponential cutoff instead of the superposition of a power law and a heavy dark matter contribution provides a comparably good fit to the HESS data with $\chi^2/\mbox{d.o.f.} = 0.78$.

\begin{figure*}[t]

\begin{tabular}{ccc}
\includegraphics[width=0.48\textwidth]{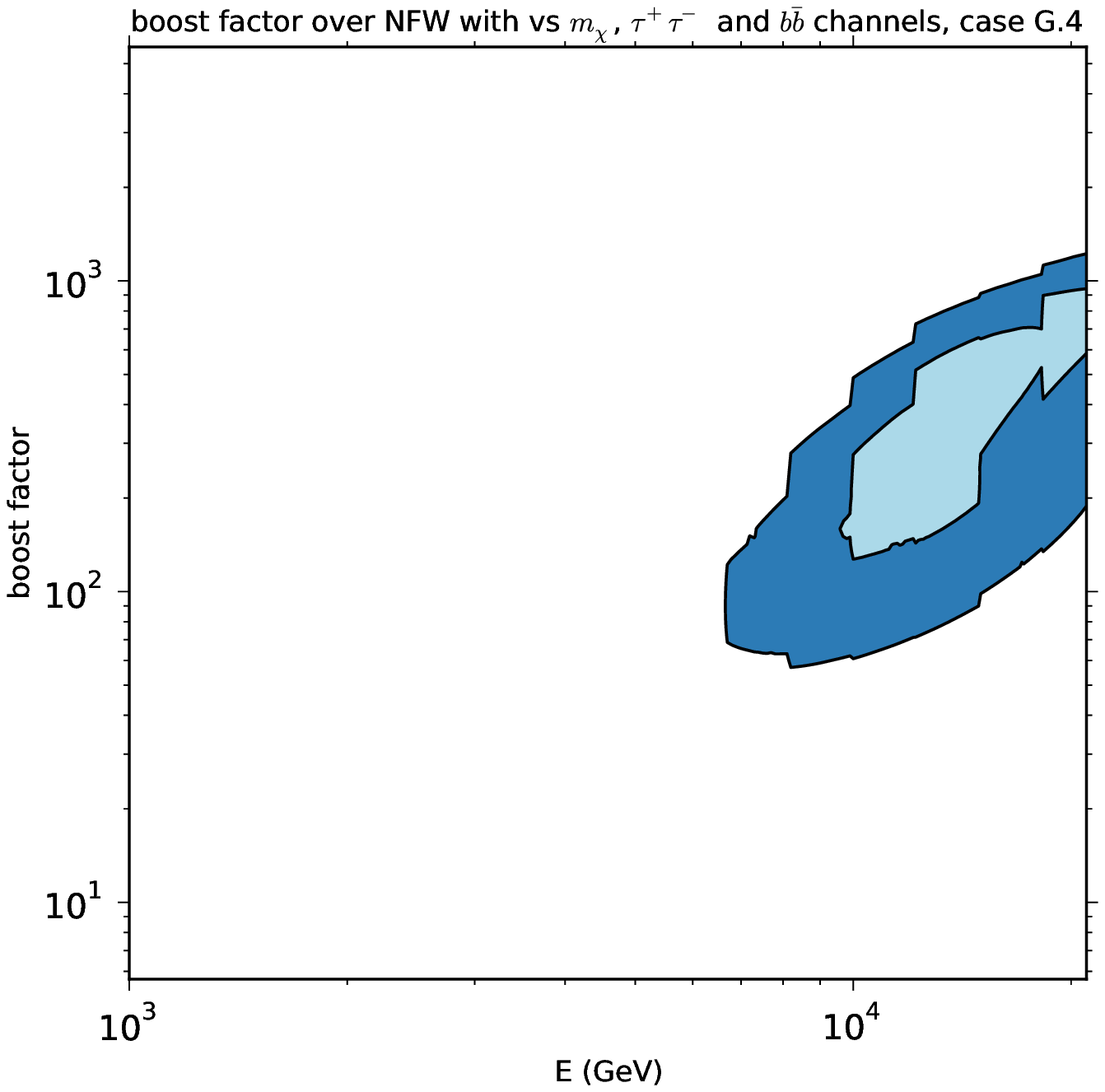} & &
\includegraphics[width=0.48\textwidth]{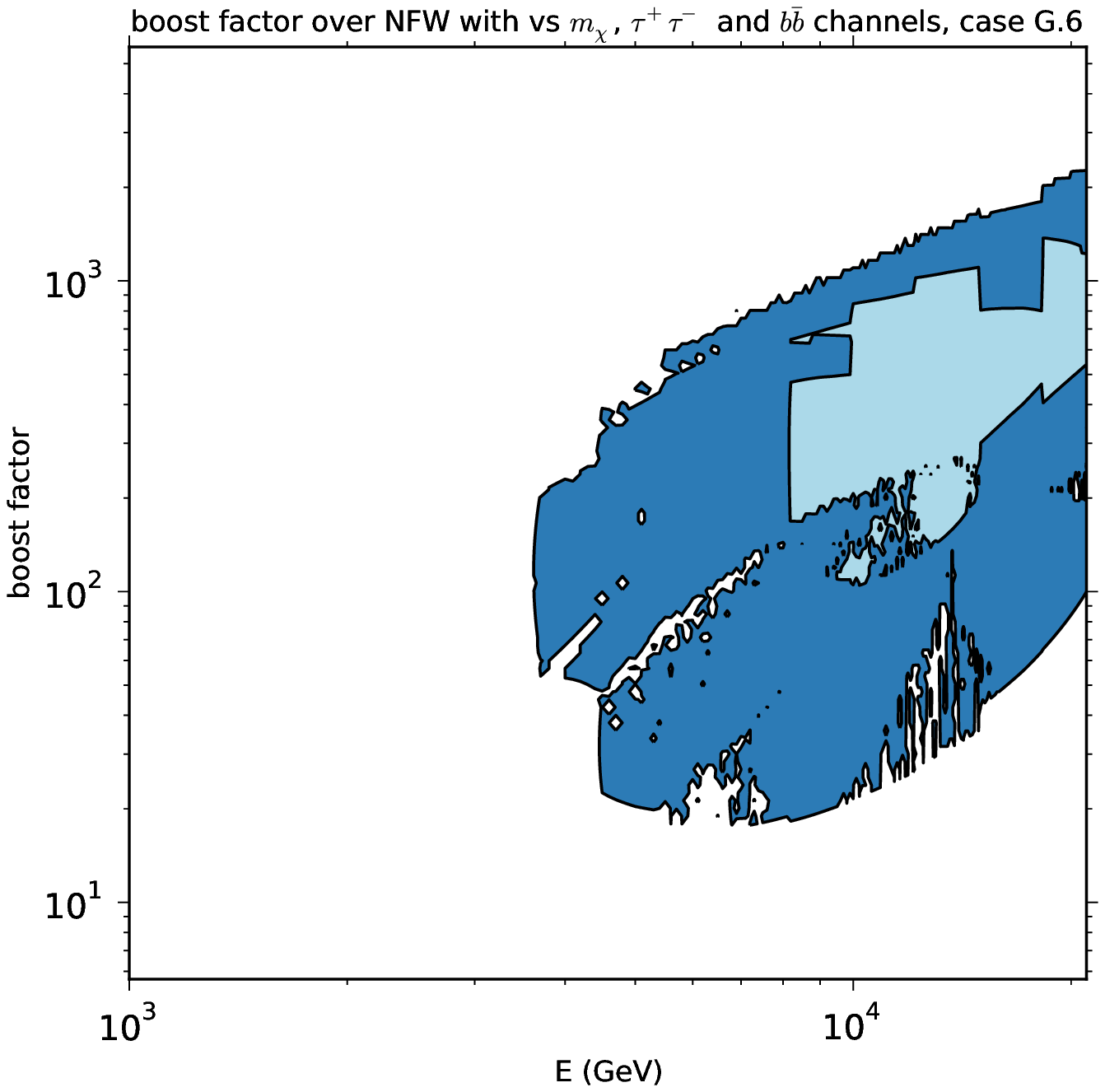} \\
&& \\
\includegraphics[width=0.48\textwidth]{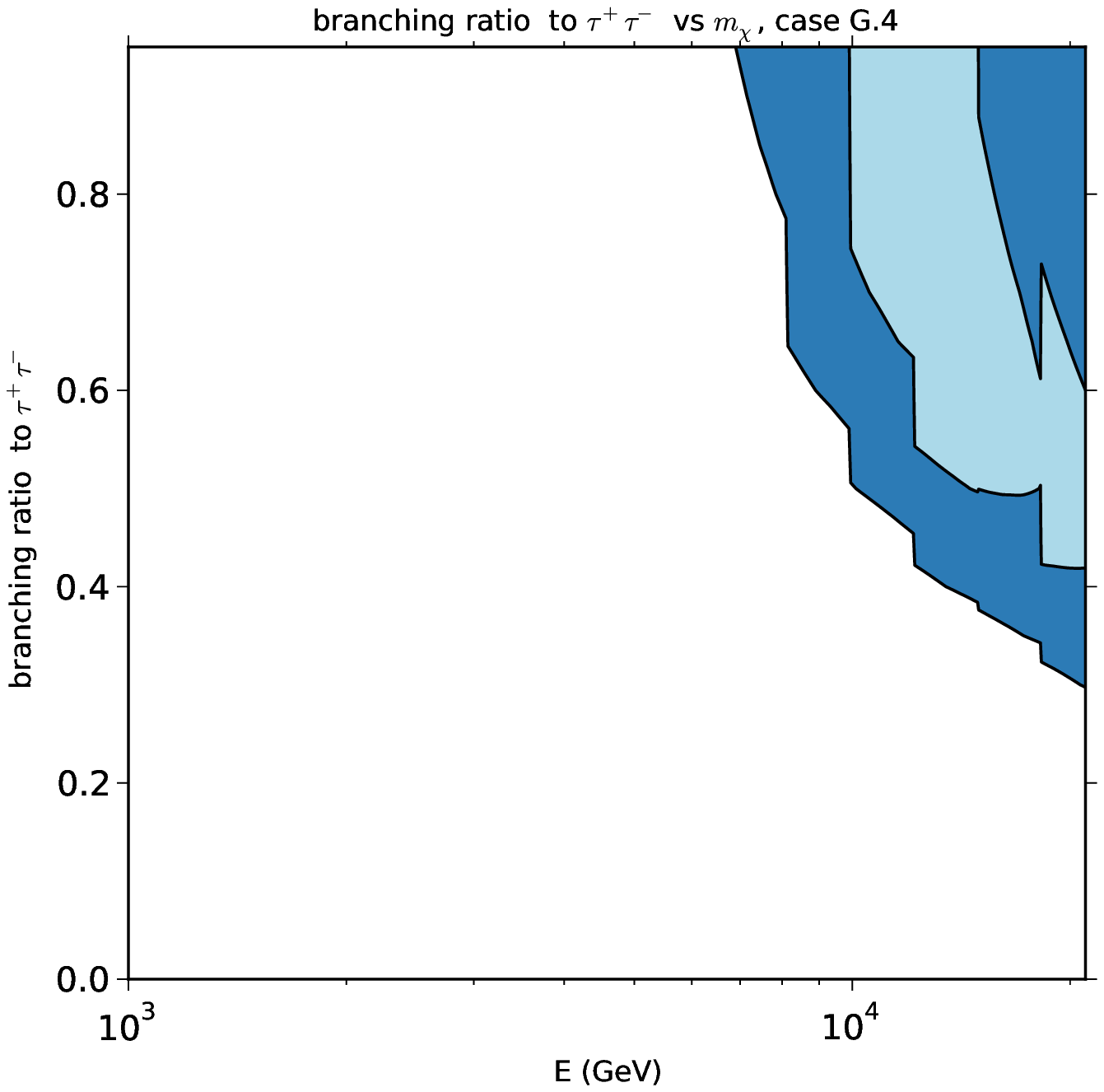} & &
\includegraphics[width=0.48\textwidth]{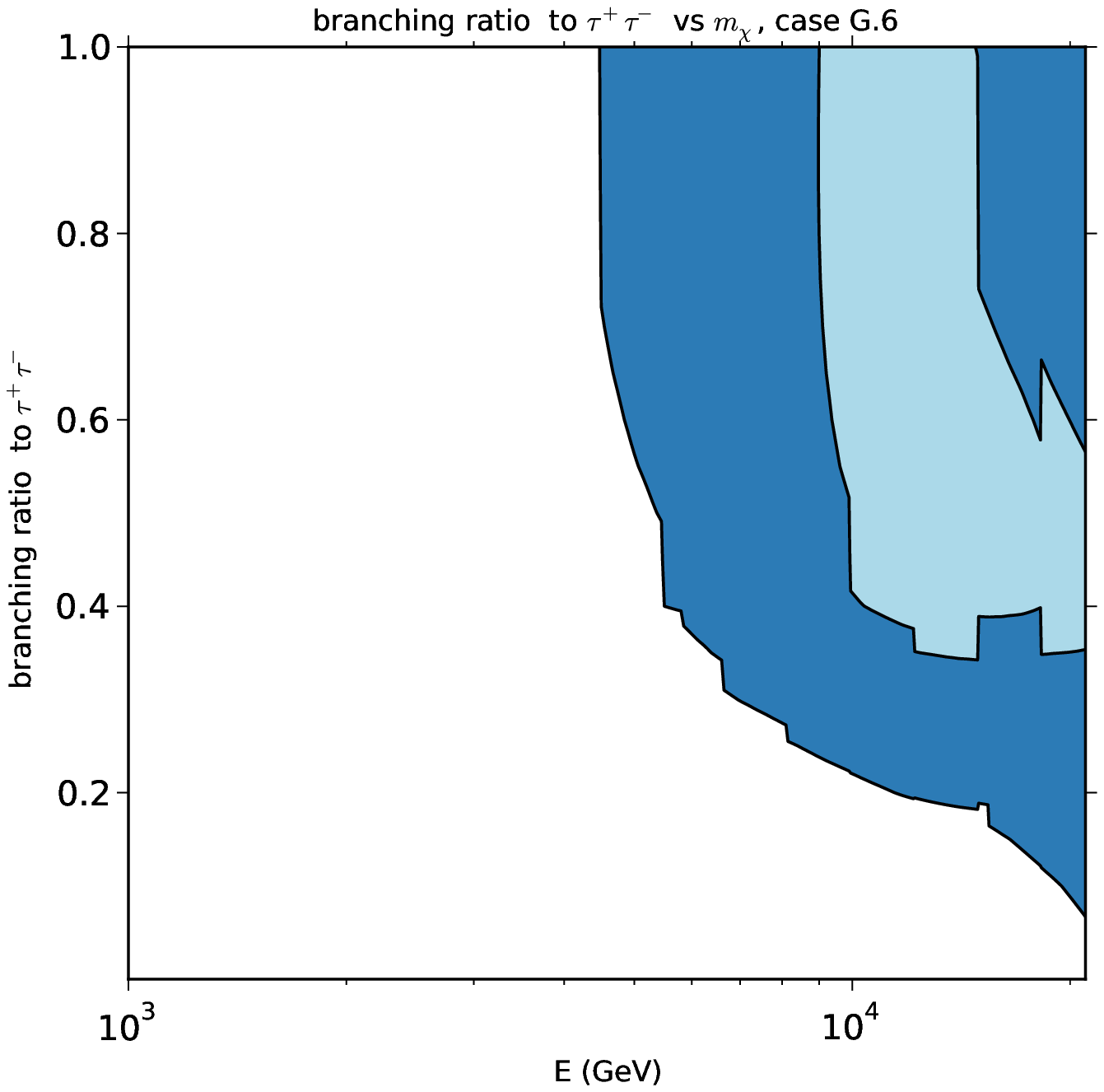} 
\end{tabular}
\caption{{\it Top panels}: allowed regions of parameter space $m_\chi$-$BF$ (mass of dark matter particle - boost factor) that fit {\it Fermi} and HESS galactic center gamma-ray data, case G.4 ({\it top left panel}) and that fit HESS data alone, case G.6 ({\it top right panel}) (the values of $\chi^2$ for given values of $m_\chi$ and $BF$ are found by minimizing over a combination of $\tau^+\tau^-$ and $b \bar b$ dark annihilation channels). 
{\it Bottom panels}: Allowed regions of parameter space $m_\chi$-$Br_{\tau^+\tau^-}$ (mass of dark matter particle - branching ratio to $\tau^+\tau^-$) that fit {\it Fermi} galactic center gamma-ray data and HESS galactic center gamma-ray data, case G.4 ({\it bottom left panel}) and the HESS data alone, case G.6 ({\it bottom right panel}). In all cases, light-blue and blue correspond respectively to 95\% and 99.999\% C.L. regions.}
\label{GC_single}
\end{figure*}
In the top panels of Fig. \ref{GC_single} we plot the regions in the parameter space $m_\chi$-$B_F$, as defined in Section~\ref{sec:dist}, for the case G.4, marginalized over all DM annihilation channels. These results are very similar to the results of case G.5. If we fit only the HESS data, the region gets somewhat enlarged towards smaller boost factor values. In the lower panels we plot instead the allowed regions of parameter space considering the DM mass versus the branching ratio to $\tau^+\tau^-$, that is $m_\chi$-$Br_{\tau^+\tau^-}$. We assume the remaining annihilation products to proceed through softer $b\,\bar b$-like channels. In both figures, light-blue and blue regions correspond respectively to 95\% and 99.999\% C.L. regions.

\section{Local population of electrons and positrons and DM spectral fits} 
\label{sec:elpos}

The standard established mechanism for producing cosmic ray (CR) positrons is by means of secondary production in the interactions of CR nuclei with interstellar gas. Their measurement was therefore recognized as a way to test models of the production and propagation of cosmic rays in the Galaxy, as well as models of new physics, in particular the annihilations or decays of dark matter \cite{Silk:1984zy}. 

This approach proved to be fruitful as the PAMELA satellite reported a surprising rise of the positron fraction $e^+/(e^+ + e^-)$ above the expected declining astrophysical background from $\sim 10$ GeV up to at least 100 GeV \cite{Adriani:2008zr} (these data confirmed the results from the earlier HEAT balloon experiment \cite{Barwick:1997ig} and AMS test-flight \cite{Aguilar:2007yf}). Secondary production mechanism results in a positron fraction that decreases with energy, once diffusion effects are taken into account \cite{Serpico:2011wg}. The rising positron fraction, also recently confirmed by {\it Fermi} LAT measurement, \cite{FermiLAT:2011ab}, therefore points to an existence of a source of primary positrons.  

At the same time, PAMELA measured anti-proton fraction, ${\bar p}/(p+\bar p)$, consistent with expectations based on astrophysical secondaries, up to the maximal probed energy of about 200 GeV \cite{Adriani:2010rc}, in agreement with earlier data. 

Consistently with the positron rise, the measurement of the total electron ($e^+ + e^-$) spectrum by {\it Fermi} LAT \cite{Abdo:2009zk} showed a hard $E^{-2}$ power law, opening the possibility that an additional hard spectral component contributes on top of a softer astrophysical spectrum. This measurement was later confirmed by PAMELA \cite{Adriani:2011xv} and updated by {\it Fermi} LAT \cite{FermiLAT:2011ab}.
The HESS telescope also reports the measurement of the $e^+ + e^-$ energy spectrum above energies of 600 GeV, showing a power law spectrum in agreement with the one from {\it Fermi} LAT and measuring a cut-off at around 3 TeV, \cite{Egberts:2011zz}. This spectrum was subsequently confirmed by the measurement by the MAGIC telescope \cite{BorlaTridon:2011dk}.

There are several candidates for astrophysical sources which may supply this extra-yield of cosmic positrons: i) pulsars could provide sizable contributions to the positron flux from pair conversions of $\gamma$ rays in their magnetic fields (for recent work see \cite{Hooper:2008kg,Yuksel:2008rf,Profumo:2008ms,Malyshev:2009tw}), ii) spallation processes  of cosmic rays inside SNRs with the gas, during the acceleration stage \cite{Blasi:2009bd,Mertsch:2009ph,Ahlers:2009ae} or iii) a more refined treatment of a spatial distribution of sources and gas  \cite{Shaviv:2009bu}. In analogy, dark matter annihilation (or decay) can also provide a needed source of positrons, provided that its annihilation cross section is enhanced with respect to the value expected for WIMPs and that dark matter is either heavy, or has suppressed decay to hadronic states, otherwise the antiproton measurements would be exceeded \cite{Donato:2008jk,Cirelli:2008pk}. 

In the comprehensive work of \cite{Delahaye:2010ji}, the authors modeled electron/positron contribution of pulsars and local supernovae, and showed that, while one cannot predict the local electron and positron fluxes (as the overall spectrum is mostly set by a hierarchy in the local sources), it is still possible to explain current measurements by means of reasonable parameterizations of the source and propagation, without requiring exotic contributions.

In early calculations of the dark matter electron/positron contribution to the locally measured spectra, energy losses were calculated assuming the Thompson regime \cite{Baltz:1998xv,Hooper:2004xn,Lavalle:1900wn,Asano:2006nr,Bergstrom:2009fa,Cirelli:2008pk,Delahaye:2007fr,Pieri:2009je,Catena:2009tm}.
However, it has been noted that the Thomson approximation is no longer valid for energies at Earth above a few tens of GeV, for which a full relativistic description is consequently necessary, \cite{Stawarz:2009ig}. 
Here we employ a full relativistic Klein-Nishina calculation as implemented in the \texttt{GALPROP} code (see also \cite{Finkbeiner:2010sm}). We also use an injection spectrum of dark matter annihilation products which takes into account electroweak bremstrahlung \cite{Cirelli:2010xx}, important for dark matter and specially for positron fraction at lower energies. 

In what follows, we set aside the question of the normalization of DM signals in the GC and local regions and ask the question  whether there is a spectral overlap between the gamma-ray feature and the feature in the electron and positron data at TeV energies. 

The data on the total flux of electrons and positrons are available in the range 1 GeV through 4.5 TeV, with PAMELA covering the lower part from 1 through 625 GeV \cite{Adriani:2011xv}, {\it Fermi} LAT in the range 20 GeV-1 TeV \cite{Abdo:2009zk} and HESS \cite{Egberts:2011zz} in the range 370 GeV - 4.5 TeV. To account for systematic uncertainties to the statistical errors provided in the HESS dataset, we have added a systematic uncertainty from the HESS band bin by bin.

We will use the term `electrons' to imply the total electron plus positron flux which we describe by the sum of a power-law astrophysical component (modulated by the solar potential when required) plus a dark matter contribution. We note that a simple power-law is not expected to necessarily be the best fit for the astrophysical electron spectrum in the high energy range, where the stochastic nature of nearby sources needs to be taken into account but we resort to this parametrization for simplicity. In addition, to circumvent this problem while exploring the position of the energy (or DM mass) cut-off, we focus only on the HESS data.

\begin{figure*}[t]
\includegraphics[width=0.49\textwidth]{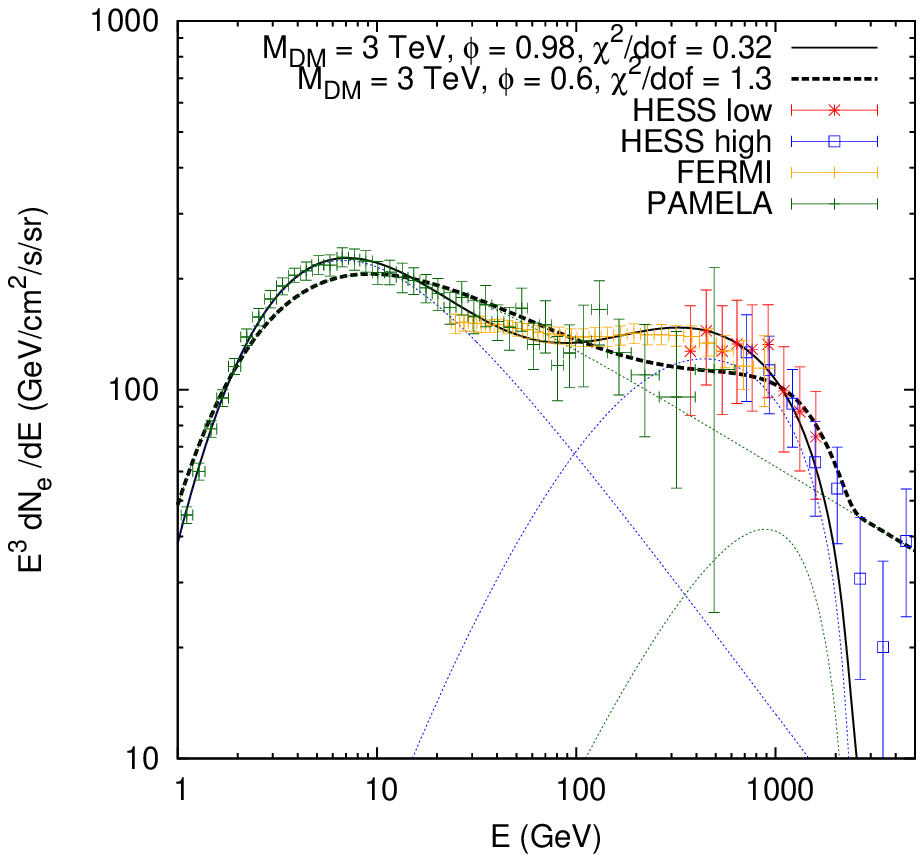}
\includegraphics[width=0.49\textwidth]{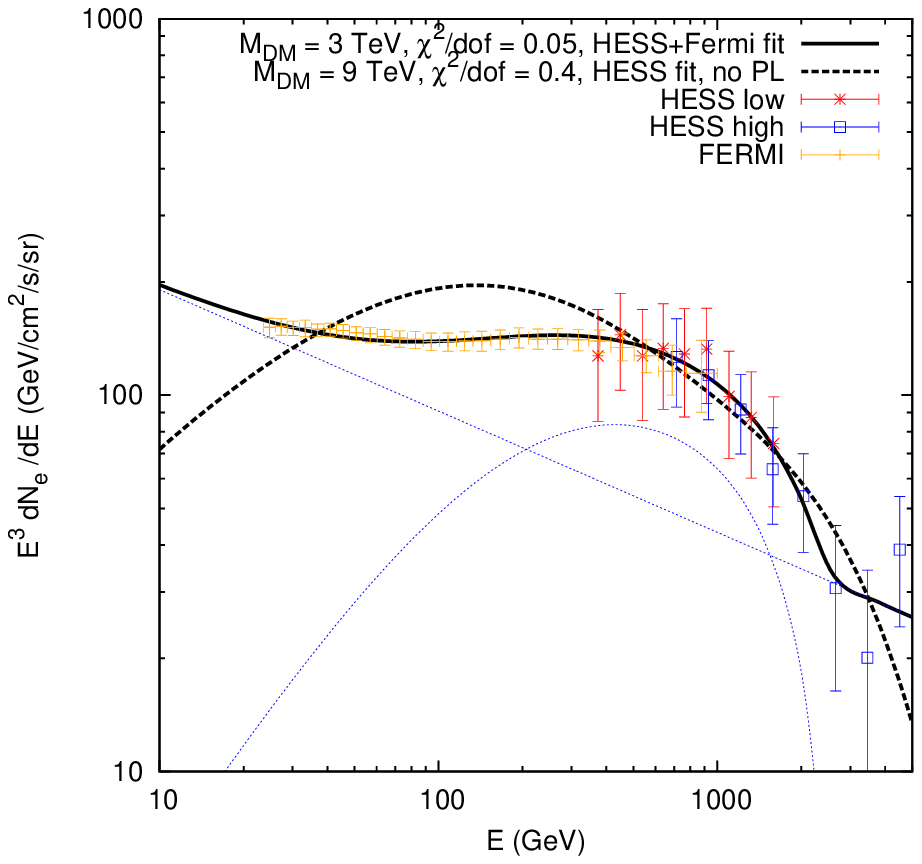}
\caption{{\it Left panel}: best fits for case E.1 ({\it solid curve}) and E.2 ({\it dashed curve}). {\it Right panel}: best fits for case E.3 ({\it solid curve}) and E.4 ({\it dashed curve}). In both panels, individual components of the fits (power law and the dark matter spectrum) are shown by the dotted lines. The best-fit values for the parameters are given in Table \ref{tabEl}.}
\label{PosEl_bestFits}
\end{figure*}

We have found the best fits for the combination of the PAMELA, {\it Fermi} LAT and complete HESS datasets  modeling the attenuation of the flux at energies below 10 GeV due to solar modulation by rescaling the spectrum \cite{Gleeson:1968zz, Mertsch:2010qf}. A first approach consists in keeping the solar modulation potential $\phi$ as a free parameter (case E.1), while a second approach assumes a fixed value $\phi = 600$ MV (case E.2). To minimize the effect of solar modulation, we have also considered a power law and dark matter fit for the {\it Fermi} LAT and the complete HESS datasets alone, thereby restricting ourselves to the range above 20 GeV (case E.3) but including a fit to the positrons data from {\it Fermi} LAT. Finally, case E.4 corresponds to fitting just the HESS datasets with dark matter. In Table~\ref{tabEl} we list the parameters values corresponding to the best fits for all mentioned cases. The results for the cases mentioned above are shown in Fig.~\ref{PosEl_bestFits}.
\begin{center}
{\begin{table*}[t]
\hfill{}
\begin{tabular}{c|c|c|c|c|c|c|c|c}
\hline
case & datasets & mass[TeV] & ($Br_{\tau^+\tau^-}$; $Br_{b\bar b}$; $Br_{\mu^+\mu^-}$) & $10^{-3}\times B_{NFW}$ & $\Gamma$ & A & $\phi$ [MV] & $\chi^2$ /d.o.f.  \\ 
\hline
E.1 & PAMELA+{\it Fermi}+HESS & 3 & (0.9997;0;0)		&	12.8	&	3.72	& 1943	& 986 & 0.3 \\
E.2 & PAMELA+{\it Fermi}+HESS & 3 & (0;0;0.9996)		&	1.38	&	3.33	& 595		& 600 & 1.3 \\
E.3 & {\it Fermi} +HESS (+{\it Fermi} $e^+$) 				& 3 & (0.9997;0;0) 		&	8.7		&	3.32	& 400		& - 	& 0.05 \\
E.4 & HESS 							& 9 & (0.035;0.965;0) &	94.9	&	-			& -			& - 	& 0.4 \\ 
\hline
\end{tabular}
\hfill{}
\caption{Fit to the electron and positron spectra \label{tab:El}}
\label{tabEl}
\end{table*}}
\end{center}

\begin{figure*}[t]
\includegraphics[width=0.49\textwidth]{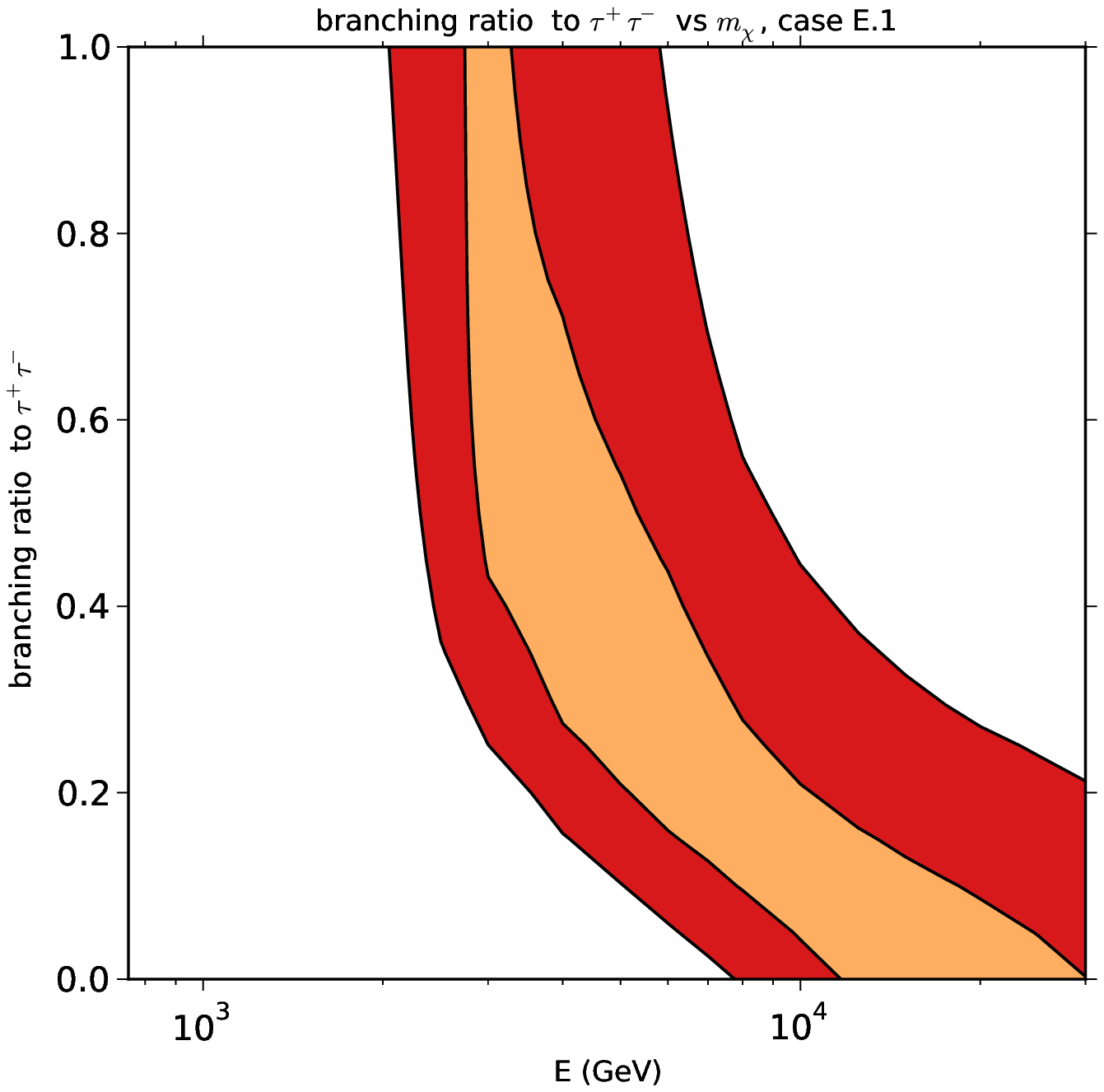}
\includegraphics[width=0.49\textwidth]{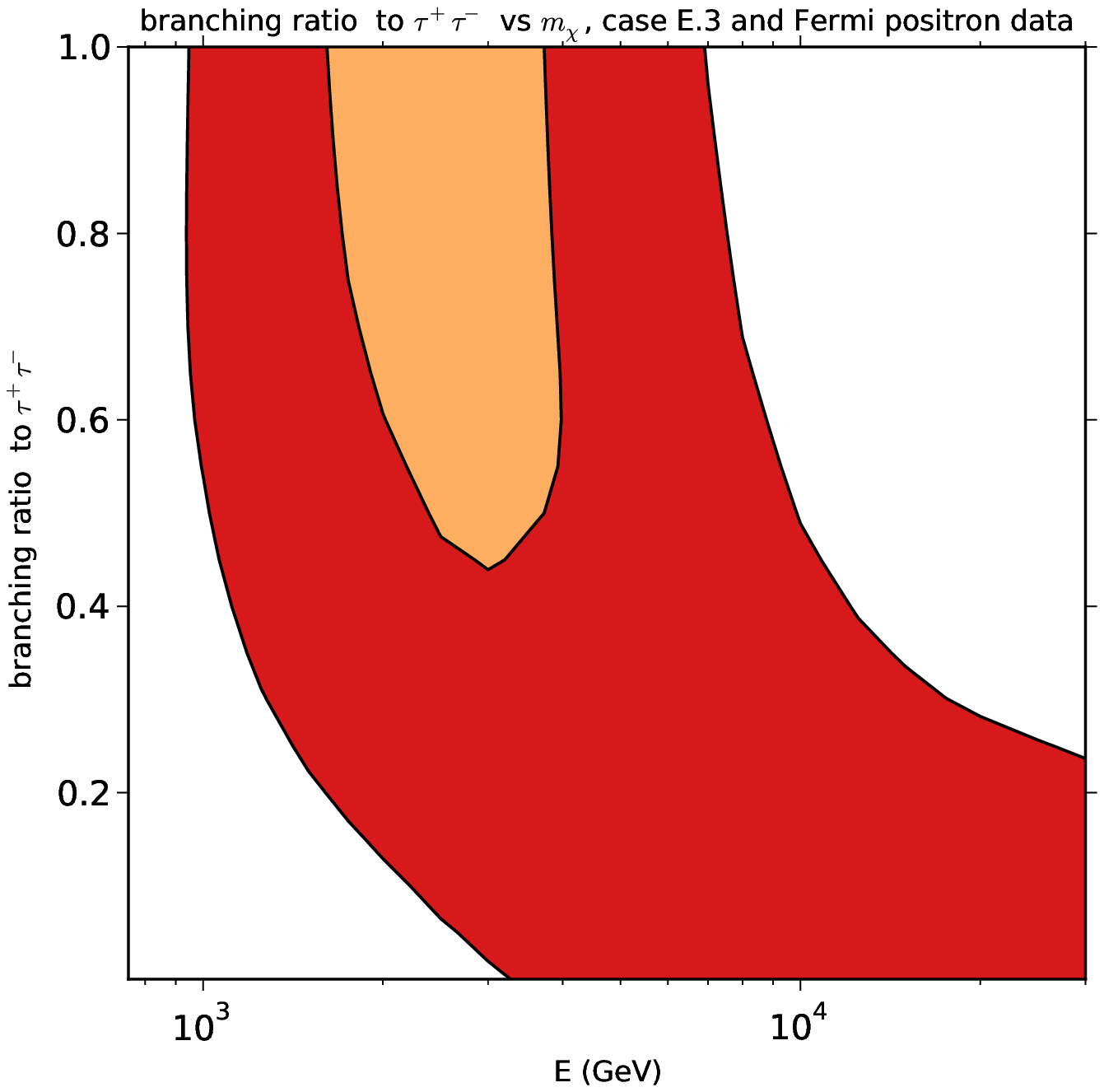}
\caption{Allowed regions of parameters space $m_\chi$-$Br_{\tau^+\tau^-}$ (DM mass - branching ratio to $\tau^+\tau^-$) from the analysis of PAMELA, {\it Fermi} LAT and HESS $e^++e^-$ data, case E.1 ({\it left panel}) and  {\it Fermi} LAT and HESS $e^++e^-$ data with, in addition, the {\it Fermi} LAT $e^+$ data, case E.3 ({\it right panel}). Orange and red correspond respectively to the 95\% and 99.999\% C.L. regions.}
\label{el_elNoPam}
\end{figure*}
Fig.~\ref{el_elNoPam} shows the regions of parameters space $m_\chi$-$Br_{\tau^+\tau^-}$ (dark matter mass - branching ratio to $\tau^+\tau^-$) that fit PAMELA, {\it Fermi} LAT and HESS electrons and positrons data with varying solar potential $\phi$, case E.1 ({\it left panel}) and $e^++e^-$ data with, in addition, the {\it Fermi} LAT $e^+$ data, case E.3 ({\it right panel}). Light-orange and red correspond respectively to the 95\% and 99.999\% C.L. regions. 

\begin{figure}[t!]
\includegraphics[width=0.49\textwidth]{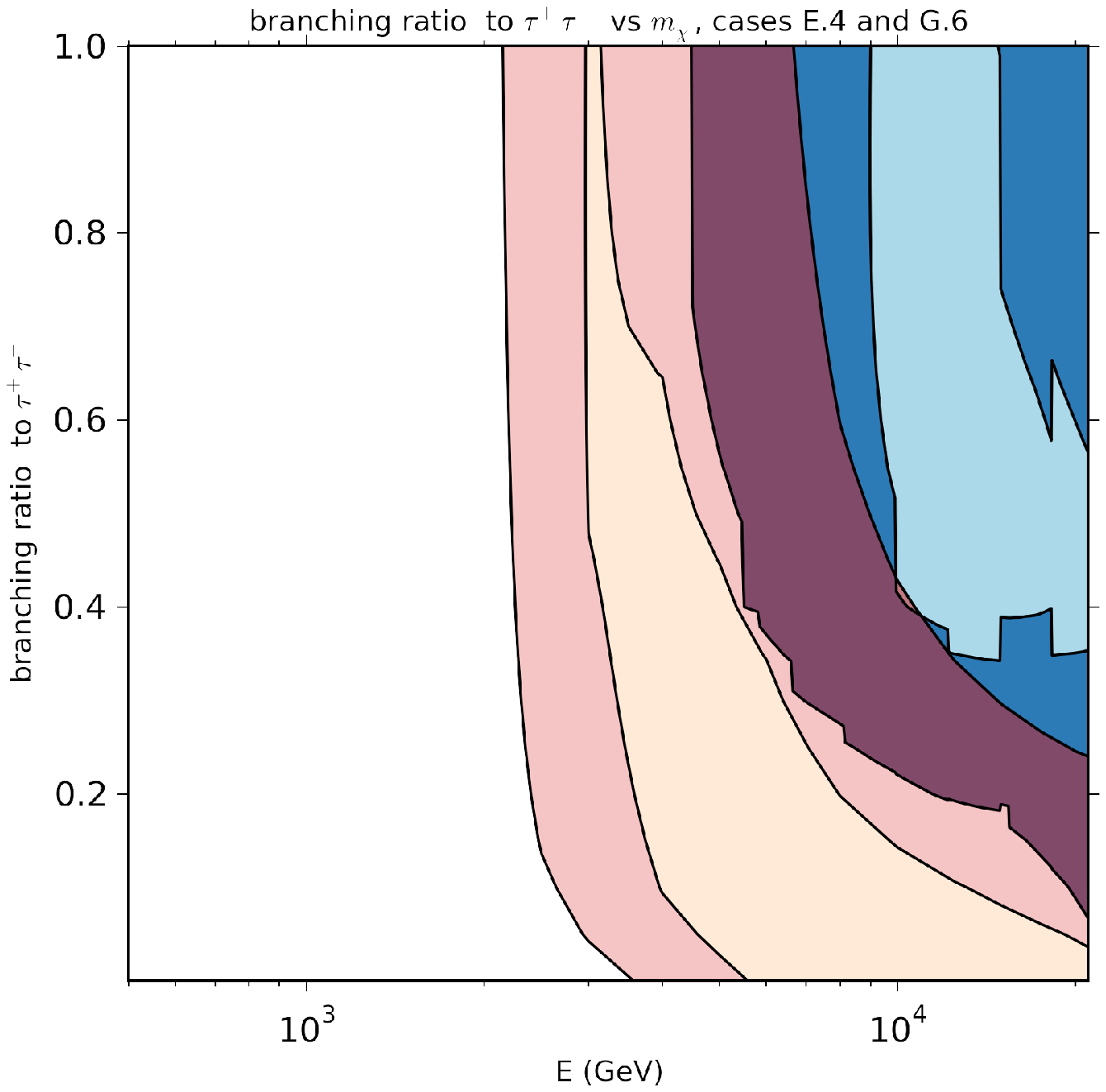}
\caption{Allowed regions of parameter space $m_\chi$-$Br_{\tau^+\tau^-}$ (DM mass - branching ratio to $\tau^+\tau^-$) from the fit of the HESS GC gamma-ray data, case G.6 and the HESS $e^++e^-$ data, case E.4. Light-blue and blue correspond to the 95\% and 99.999\% C.L. regions from the gamma-ray data and orange and red correspond instead to the 95\% and 99.999\% C.L. regions from the electron data.}
\label{elPhi_GC}
\end{figure}
In Fig. \ref{elPhi_GC}, we plot the allowed regions in the same parameter space $m_\chi$-$Br_{\tau^+\tau^-}$ from the HESS GC gamma-rays analysis, case G.6, and HESS $e^++e^-$ data, case E.4. 

Finally, Fig. \ref{El_single} shows the regions in the parameters space $m_\chi$-boost factor for the $e^++e^-$ data corresponding to case E.3. The boost factors required by the DM fits in this case are an order of magnitude higher than the ones implied by the GC gamma-ray data. This poses an additional difficulty in relating these two phenomena and at the same time points to the fact that a potential signal in the GC regions would not be in a conflict with the local measurement of electrons/positrons.

In order to alleviate the problem of high boost factors, an explanation of the electron flux in terms of dark matter annihilating in a nearby clump has been proposed in \cite{Hooper:2008kv, Regis:2009qt}. The normalization of the flux then depends on the size and the density profile of the clump as well as on the distance to the clump. The farther away we place the clump, the bigger and/or denser we have to make it to achieve the same required boost factor.
The effect on the spectral shape, however, is not degenerate.
In fact, the high-energy cutoff in the electron spectrum is inversely proportional to the distance to the source of electrons \cite{Kobayashi:2003kp,Egberts:2011zz}.
In such a way a spectrum of electrons which initially extended to higher energies would appear to us as being cut off to lower energies and it is conceivable that an explanation of electron flux in terms dark matter annihilating in a nearby clump could be more spectrally consistent with the explanation of gamma-ray data from the galactic center. 

However, large clumps needed are very unlikely at the required distance to the Solar neighborhood \cite{Springel:2008cc} and it was shown in  \cite{Lavalle:1900wn} that the average contribution of the clumps is expected to be negligible compared to that of the smooth dark matter component. Despite these considerable difficulties, a study of electrons due to clumpy DM in the light of models we consider here could be of potential interest.


\section{Constraints on annihilation cross-section for heavy DM candidates} \label{sec:const}

In this section, we briefly compare our results with current constraints, focusing on heavy ($\gsi 1$ TeV) DM candidates (for a recent review on indirects constraints on DM annihilation cross section see \cite{Cirelli:2012tf}.) We should underline that the analysis in this work relies mainly on spectral features, whose normalization constrain the product of cross section and dark matter density. The significant uncertainty on the density in this particular region results in turn in a large uncertainty on our predictions for the annihilation cross section ($\langle \sigma v \rangle$) so that comparisons with the limits on  $\langle \sigma v \rangle$ listed below should therefore be taken only approximately.

\begin{itemize}

\item {\em HESS GC halo analysis} \cite{Abramowski:2011hc} (see also \cite{Abazajian:2011ak}). HESS  observations of the Galactic Center present the strongest constraints on WIMP dark matter annihilation into Standard Model particles for $\gsi 1$ TeV, given a dark matter Milky Way density described by a non-adiabatically-contracted NFW or Einasto profile. DM signal search in  \cite{Abramowski:2011hc}  is performed with a projected galactocentric distance of 45 pc -- 150 pc (corresponding to an angular distance of $0.3^\circ$ -- $1.0^\circ$), excluding the Galactic plane. Note, however, that for an isothermal halo profile, the signal would be completely subtracted in this analysis (as backgrounds are modeled according to the signal in the sorrounding region) and therefore if the DM signal is flat at distances $\gsi 45$ pc, this limit does not apply. Limits are derived under the assumption that DM particles annihilate into quark-antiquark pairs, and constraints DM annihilation cross sections of the order $2\times10^{-24}$ cm$^3$s$^{-2}$, or boost factors of ~140 in our notation (when we renormalize to the same local density of DM) at 10 TeV. 

\item {\it GC ridge}  \cite{Aharonian:2006au}. A band of diffuse emission along the GC ridge, was used to derive upper limits on a DM induced signal \cite{Crocker:2010gy,Bertone:2008xr,Meade:2009iu}. The limits are somewhat worse than in the analysis mentioned above, and are of the order of $10^{-23}$ cm$^3$s$^{-2}$, or boost factors of ~700 in our notation at 10 TeV, for the annihilation channel to $\tau ^+\tau^-$.

\item  {\it PAMELA antiproton data} \cite{Adriani:2010rc} were used to set limits on DM annihilations to hadronic final states \cite{Donato:2008jk,Cirelli:2008pk}. Recently \cite{Garny:2011cj,Garny:2011ii} considered the case in which the dark matter is constituted by Majorana particles and studied the annihilation process into two fermions and one gauge boson which, under some circumstances, can have a non-negligible or even a larger cross section than the lowest order annihilation process into two fermions as we discussed in section \ref{sec:spectrum}. They performed a general analysis of this scenario, calculating the annihilation cross section of the three body final state processes, $\chi \chi \rightarrow f {\bar f} V$ when the dark matter particle is a $SU(2)_L$ singlet or doublet, where $f$ is a lepton or a quark, and $V$ is a photon, a weak gauge boson or a gluon. This framework is relevant in the context of this analysis as heavier DM is expected to produce an additional gauge boson more readily.They set a limit on the value of the  boost factor at $\sim10^3$ at 5 TeV, for the case of singlet DM coupled to quarks (limits in the case of a coupling to leptons are weaker).

\item {\it Neutrino observations} (ICE CUBE \cite{Abbasi:2011eq} and SuperKamiokande \cite{Desai:2007ra,Meade:2009iu}). Such observations have the advantage that higher energy neutrinos coming from heavier DM annihilations also have a higher cross section for detection and thus the loss in rate is partially compensated by the increase in sensitivity. Neutrino constraints become therefore somewhat competitive with $\gamma$-ray bounds at large DM masses, reaching cross sections of $\sim10^{-22}$ cm$^3$s$^{-2}$ at 10 TeV.

\begin{figure}[t!]
\includegraphics[width=0.53\textwidth]{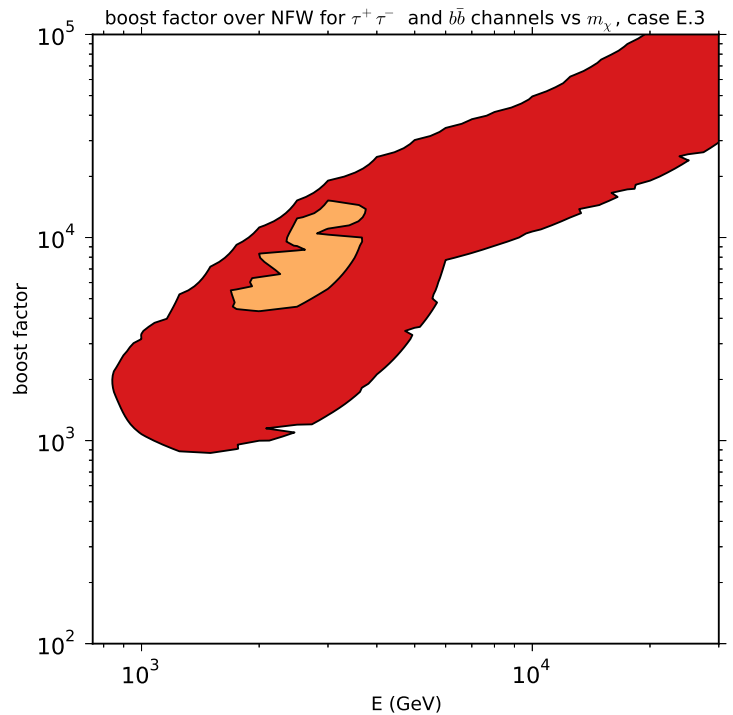}
\caption{The regions of parameter space $m_\chi$-BF (mass of dark matter particle - boost factor) that fit {\it Fermi} and HESS electron data, case E.3 (the values of $\chi^2$ for given values of $m_\chi$ and $BF$ are found by minimizing over a combination of $\tau^+\tau^-$ and $b \bar b$ dark annihilation channels). Faded orange and red correspond to 95\% and 99.999\% C.L. regions.}
\label{El_single}
\end{figure}

\item  {\it Multi-wavelength analysis}. A systematic study of the multi--wavelength signal induced by WIMP annihilations at the GC was performed in \cite{Regis:2008ij} (see also \cite{Bergstrom:2006ny} for X-ray constraints, and \cite{Gondolo:2000pn,Bertone:2001jv}, for synchrotron constraints). The authors shown that the luminosity of the WIMP source and experimental sensitivity at different frequencies is at a comparable level, depending rather weakly on WIMP mass and annihilation channels. For 10 TeV DM mass and $\tau^+\tau^-$ annihilation channels the most stringent constraints come from the near-infrared VLT data in the Ks band (2.16 $\mu$m) based on the observation within the central 10-20 milliarcsec  ($\lsi 0.01$ pc). For the assumed {\it adiabatically contracted} DM profile (enhanced by about a factor of 100 with respect to the NFW profile in that region) limits are at the level of $\langle\sigma v\rangle\lsi 10^{-26}$ cm$^{-3}$s$^{-1}$, roughly consistent with the ones derived in this work. These limits however depend on the details of the DM density profile as well as on the energy loss and propagation parameters of electrons produced in DM annihilations in the close vicinity of the black hole.

\item {\it Unitarity arguments}. In \cite{Griest:1989wd} it was shown that unitarity arguments imply interesting constraints on  strongly annihilating dark matter constraining annihilation cross sections to be weaker than $\sim10^{-21}$ cm$^3$s$^{-2}$ at 10 TeV.

\end{itemize}

\section{Summary} \label{sec:summary}

In this work,  we fit the gamma ray spectrum of the central Galactic source over five decades in energy from the joint {\it Fermi} LAT and HESS data by making several choices to model the astrophysical emission and adding a heavy DM contribution. In particular we model the joint spectra with a combination of a power law, a  spectrum of unresolved pulsars and a DM contribution. Good fits are achieved with a power law with a  slope index of $\sim 2.5-2.6$, an exponential cut-off with index similar to those of the observed pulsar population and $\gsi10$ TeV DM decaying to a mixture of $b\,{\bar b}$ and harder $\tau^+\, \tau^-$ channels. We show that, alternatively, the GC gamma-ray spectrum can be well fit with a log-parabola and  heavy DM spectrum alone. The best fit in this case is realized by a somewhat heavier DM annihilating with higher branching ratios to channels resulting in softer f $b\,{\bar b}$-like spectra. We also performed a fit to the HESS data alone, and conclude that the DM parameter space is less constrained in this case, allowing for a wider mass range and combinations of hard and soft channels. 

The best fit DM models generally require boost factors of the order of 100-300 (for a DM mass of 10 TeV), when compared to the values expected based on the NFW profile and a thermal annihilation cross-section. These values are consistent or in a mild tension with the constraints derived analyzing the HESS data in the region 45-100 pc away from the GC  \cite{Abramowski:2011hc}. However, the DM profiles in the vicinity of the super massive black hole could be slightly enhanced with respect to their shape at distances larger than $50$ pc away, where GC halo constraints were derived, thereby loosening such constraints.

In addition, by leaving a large freedom in our parameters, we explore whether there might be an overlap in the DM parameter space obtained when considering the Galactic center gamma ray and local electron/positron data. We studied the DM mass-branching ratio to $\tau$ allowed parameter space, $m_\chi$-$Br_{\tau^+\tau^-}$, for the gamma ray and electrons and positrons data, based on spectral features and find that the 99.999\% C.L. regions overlap only marginally.

\section*{Acknowledgments}
We thank Kathrin Egberts and Daniela Borla Tridon for providing or referring to the data. We kindly thank Torsten Bringmann, Illias Cholis, Marco Cirelli, Emanuel Moulin, Emiliano Sefusatti, Pasquale Serpico and Andrea de Simone  for numerous discussions. AB was supported by a grant of 47th Rencontres de Moriond meeting. GZ is grateful to the Institut d'Astrophysique de Paris for hospitality during completion of this project. 

\bibliographystyle{unsrt}	
\bibliography{gc_final_Arxiv}		

\end{document}